\documentstyle[psfig]{jfm}
\def\be{\begin{equation}}
\def\ee{\end{equation}}

\begin{document}

\title 
[Exact solution of the Riemann problem in RMHD with tangential
magnetic fields]
{The exact solution of the Riemann problem in relativistic MHD with
tangential magnetic fields}

\author
[R. Romero et al.]
{Roberto Romero$^1$, Jos\'e M$^{\underline{\mbox a}}$ Mart\'{\i}$^1$,
Jos\'e A. Pons$^2$, \\ 
Jos\'e M$^{\underline{\mbox a}}$ Ib\'a\~nez$^1$
and Juan A. Miralles$^2$}
              
\affiliation{$^1$Departament d'Astronomia i Astrof\'{\i}sica \\
Universitat de Val\`encia, 46100 Burjassot, Spain \\[\affilskip]
$^2$ Departament de F\'{\i}sica Aplicada \\
Universitat d'Alacant, Ap. de correus 99, 03080 Alacant Spain}

\date{\today}

\maketitle

\begin{abstract}

  We have extended the procedure to find the {\it exact} solution of
the Riemann problem in relativistic hydrodynamics to a particular case
of relativistic magnetohydrodynamics in which the magnetic field of
the initial states is tangential to the discontinuity and orthogonal
to the flow velocity. The wave pattern produced after the break up of
the initial discontinuity is analogous to the non--magnetic case and
we show that the problem can be understood as a purely relativistic
hydrodynamical problem with a modified equation of state. The new
degree of freedom introduced by the non-zero component of the magnetic
field results in interesting effects consisting in the change of the
wave patterns for given initial thermodynamical states, in a similar
way to the effects arising from the introduction of tangential
velocities. Secondly, when the magnetic field dominates the
thermodynamical pressure and energy, the wave speeds approach the
speed of light leading to fast shocks and fast and arbitrarily thin
rarefaction waves. Our approach is the first non-trivial {\it exact}
solution of a Riemann problem in relativistic magnetohydrodynamics and
it can also be of great interest to test numerical codes against known
analytical or exact solutions\footnote{The code computing the
exact solution is available under request to the authors. Users of the
code can give credit by mentioning the source and citing this paper.}.
\end{abstract}

%%%%%%%%%%%%%%%%%%%%%%%%%%%%%%%%%%%%%%%%%%%%%%%%%%%%%%%%%%%%%%%%%%%%%%%%%%%%%% 
\section{Introduction}
\label{s:intro}

  The decay of a discontinuity separating two constant initial states
({\it Riemann problem}) has played a very important role in the
development of numerical codes for classical (Newtonian) hydrodynamics
after the pioneering work of Godunov (1959). Nowadays, most modern
high-resolution shock-capturing methods (LeVeque 1992) are based on
the exact or approximate solution of Riemann problems between adjacent
numerical cells and the development of efficient Riemann solvers has
become a research field in numerical analysis in its own (see, e.g.,
Toro 1997).  The success of high-resolution shock-capturing methods in
many areas of computational fluid dynamics has triggered their
extension to classical magnetohydrodynamics (e.g., Brio \& Wu 1988;
for an up-to-date discussion of the issue, see Balsara 2004).

As in other fields in physics, during the last two decades
astrophysics, relativity and cosmology have become computational
sciences. Modeling and understanding fluid dynamics in astrophysical
scenarios is now a key part in research projects involving supernovae,
relativistic jets, neutron star instabilities, or accretion onto
compact objects, who share a common distinctive feature: either
special or general relativity effects are relevant.  With this
motivation, Riemann solvers have been introduced in numerical
relativistic hydrodynamics since the beginning of the nineties
(Mart\'{\i}, Ib\'a\~nez \& Miralles 1991). Presently, the use of
high-resolution shock-capturing methods based on Riemann solvers is
considered the best strategy to solve the equations of relativistic
hydrodynamics which has caused the rapid development of Riemann
solvers for both special and general relativistic hydrodynamics (see,
e.g., Mart\'{\i} \& M\"uller 2003, Font 2003).

  The finding by Mart\'{\i} \& M\"uller (1994) of the analytical
solution for initial states where the flow is normal to the initial
discontinuity boosted the efforts to develop exact Riemann solvers for
relativistic hydrodynamics. Pons, Mart\'{\i} \& M\"uller (2000)
extended the domain of solutions to problems with arbitrary initial
velocities. Later, Rezzolla \& Zanotti (2001), for purely normal
flow, and Rezzolla, Zanotti \& Pons (2003), for the general
case, proposed a new procedure to find the solution
of the Riemann problem that uses the relativistically invariant
relative velocity between the unperturbed initial states. However,
up to date, no analytical or exact solution of the equations of
the Riemann problem in relativistic magnetohydrodynamics has been derived.

  The equations of both classical and relativistic
magnetohydrodynamics form a non-strictly hyperbolic system.  A
consequence of the non-strict hyperbolicity of the MHD are the
degeneracies in the wave speeds (that lead, e.g., to compound waves,
admisible solutions of the planar MHD that involve intermediate
shocks), which must be handled analytically with care and hinder the
development of both exact and approximate Riemann solvers using the
characteristic information for the MHD equations. In the relativistic
case the difficulties in the development of such solvers are increased
by the higher non-linearity of the RMHD system of equations. The
characteristic structure of the equations of RMHD is analyzed in the
books by Lichnerowicz (1967) and Anile (1989). Approximate Riemann
solvers using the characteristic information have been developed by
Romero et al. (1996), for the same particular magnetohydrodynamic
configuration considered here, and by Balsara (2001), Komissarov
(1999) and Koldoba, Kuznetsov \& Ustyugova (2002), for the general
case.  A number of analytical solutions involving only shocks,
rarefactions and Alfven waves (Komissarov 1999, Komissarov 2003) have
also been recently derived.

  In this paper we describe the solution of the Riemann problem for
the particular case in which the flow speed has two non-vanishing
components and the magnetic field is orthogonal to them. Besides this,
we force the flow to have dependence on one spatial coordinate taken
along one of the two non-vanishing velocity components. For this
particular setup, the Riemann structure degenerates to only three
waves making the solution attainable. The solution reveals interesting
and distinct properties of RMHD and could serve as a guide in the way
to the general RMHD Riemann solution. 

  In this paper we closely follow the structure and notation used in
Pons, Mart\'{\i} \& M\"uller (2000; hereafter PMM). The paper is
organized as follows. Section~\,2 collects the relevant
equations. Sections~\,3 and 4 describe, respectively, the flow across
rarefactions and shocks setting the ingredients for the Riemann
solution, which is discussed in Sect.~\,5. Conclusions are gathered in
Sect.~\,6.

%%%%%%%%%%%%%%%%%%%%%%%%%%%%%%%%%%%%%%%%%%%%%%%%%%%%%%%%%%%%%%%%%%%%%%

\section{Equations} 
\label{s:eqs}

  Let $J^{\mu}$, $T^{\mu \nu}$ and $F^{*\mu \nu}$ ($\mu,
\nu=0,1,2,3$) be the components of the density current, the
energy--momentum tensor and the Maxwell dual tensor of an ideal
magneto-fluid, respectively
\begin{equation}
J^\mu = \rho u^\mu
\end{equation}
\begin{equation}
T^{\mu \nu}= \rho \hat{h} u^\mu u^\nu +\eta^{\mu \nu}\hat{p}-b^\mu b^\nu
\end{equation}
\begin{equation}
F^{*\mu \nu}=u^\mu b^\nu -u^\nu b^\mu
\end{equation}
where $\rho$ is the proper rest-mass density, $\hat{h}=1+\epsilon
+p/\rho +b^2/\rho$ is the specific enthalpy including the contribution
from the magnetic field ($b^2$ stands for $b^\mu b_\mu$), $\epsilon$
is the specific internal energy, $p$ the thermal pressure,
$\hat{p}=p+b^2/2$ the total pressure, and $\eta^{\mu
\nu}=diag(-1,1,1,1)$ the Minkowski metric in Cartesian
coordinates. Throughout the paper we use units in which the speed of
light is $c=1$.

  The four-vectors representing the fluid velocity and the magnetic
field in the fluid rest frame, $u^\mu$ and $b^\mu$, satisfy the
conditions $u^\mu u_\mu = -1$ and $u^\mu b_\mu = 0$, and there is an
equation of state $p = p(\rho, \epsilon)$ that closes the system. All
the discussion will be valid for a general equation of state but
results will be shown for an ideal gas, for which $p = (\gamma -1)\rho
\epsilon$, where $\gamma$ is the adiabatic exponent.

  The equations of ideal RMHD correspond to the conservation of
rest mass and energy-momentum, and the Maxwell equations. In flat
space-time and Cartesian coordinates, these equations read:
\begin{equation}
\label{cont}
J^\mu_{\,\,\,\,,\mu} = 0
\end{equation}
\begin{equation}
\label{e-mom}
T^{\mu \nu}_{\,\,\,\,\,\,,\mu} = 0
\end{equation}
\begin{equation}
\label{Maxwell} 
F^{*\mu \nu}_{\,\,\,\,\,\,\,\,,\mu} = 0
\end{equation}

  We consider a particular case in which the flow speed has two
components and the magnetic field is orthogonal to them. Besides this,
we force the flow to have dependence on one spatial coordinate ($x$)
taken along one of the two non-vanishing velocity components. Specifically
we set $u^\mu = W(1, v^x, 0, v^z)$, $b^\mu = (0, 0, b, 0)$, where $W$
is the flow Lorentz factor. With these restrictions, the above system
can be written as a system of conservation laws
\begin{equation}
  \frac{\partial {\bf U}}{\partial t} + \frac{\partial {\bf F}}{\partial x} = 0
\label{eq:cl}
\end{equation}
where 
\begin{equation}
{\bf U} = (D, \hat{S}^x, \hat{S}^z, \hat{\tau}, B)^{\rm T}
\end{equation}
is the state vector of conserved quantities and 
\begin{equation}
{\bf F} = (Dv^x, \hat{S}^x v^x + \hat{p}, \hat{S}^z v^x, 
\hat{S}^x, B v^x)^{\rm T}
\end{equation}
is the corresponding vector of fluxes, with
\begin{equation}
D = \rho W
\end{equation}
\begin{equation}
\hat{S}^i = \rho \hat{h} W^2 v^i, \,\,\, (i = x,z)
\end{equation}
and
\begin{equation}
\hat{\tau} = \rho \hat{h}W^2 - \hat{p}
\end{equation}
being the rest-mass, momentum and total energy densities, and
\begin{equation}
B = bW
\end{equation}
the $y$-component of the magnetic field as measured in the laboratory
frame. Hence, according to these equations, the particular initial
configuration chosen together with the imposed symmetry prevent the
generation of new components of the velocity and magnetic field.

  It is worth noting that for the particular configuration chosen, the
term $(b^\mu b^\nu)_{,\mu}$ appearing in the equation of conservation
of the stress-energy tensor, vanishes and the RMHD equations reduce to
the purely hydrodynamical case with the only contributions from the
magnetic field appearing in the pressure and specific enthalpy, and an
additional continuity equation for the evolution of the transversal
magnetic field. This fact is considered in the Appendix where we
explore the possibility of including the magnetic effects of the
present configuration in the definition of the equation of state.

  According to the previous discussion, the magnetized flow under
consideration falls in one of the two degeneracies of the RMHD system
(Degeneracy I; Komissarov 1999), for which a description in
terms of just three characteristic waves (namely the entropy wave and
the two fast magnetosonic waves) is adequate. Turning now towards the
solution of the Riemann problem in this particular case, the
discontinuity in the initial states breaks down into a couple of left
and right propagating rarefaction waves (self-similar continuous
flows) and/or shocks and a central tangential discontinuity across
which the total pressure, $\hat{p}$, is constant. Both thermal and total
pressure increase at fast magnetosonic shocks. Hence, we would use the
comparison of the total pressure at the two limiting states to select
between shocks and rarefaction waves. 

%%%%%%%%%%%%%%%%%%%%%%%%%%%%%%%%%%%%%%%%%%%%%%%%%%%%%%%%%%%%%%%%%%%%%%

\section{Flow across rarefactions}
\label{s:raref}

  Rarefaction waves are self-similar solutions of the equations that
depend on $x$ and $t$ only through the combination $\xi=x/t$. By
imposing such a dependence in system (\ref{eq:cl}) above we obtain the
following set of equations
\begin{equation}
\label{sist1}
  (v^x - \xi)\frac{d\rho}{d\xi} + \rho(1 + v^xW^2(v^x -
  \xi))\frac{dv^x}{d\xi} + \rho W^2v^z(v^x - \xi)\frac{dv^z}{d\xi} = 0
\end{equation}
\begin{equation}
\label{sist2}
  W^2\rho \hat{h}(v^x - \xi)\frac{dv^x}{d\xi} + b(1 -
  v^x\xi)\frac{db}{d\xi} + (1 - v^x\xi)\frac{dp}{d\xi} = 0
\end{equation}
\begin{equation}
\label{sist3}
  W^2\rho \hat{h}(v^x - \xi)\frac{dv^z}{d\xi} - v^zb\xi\frac{db}{d\xi} -
  v^z\xi\frac{dp}{d\xi} = 0
\end{equation}
\begin{equation}
\label{sist4}
  \frac{dp}{d\xi} = hc_s^2\frac{d\rho}{d\xi}
\end{equation}
\begin{equation}
\label{sist5}
  b(1 + v^xW^2(v^x - \xi))\frac{dv^x}{d\xi} + bW^2v^z(v^x -
  \xi)\frac{dv^z}{d\xi} + (v^x - \xi)\frac{db}{d\xi} = 0,
\end{equation}
similar to the system obtained in PMM. The quantity $c_s$ is the sound
speed, defined by
\be 
c_s =\sqrt{\frac{1}{h}\left.\frac{\partial p}{\partial \rho}\right|_s} 
\ee 
where $s$ is the specific entropy and $h = 1 + \varepsilon +
  p/\rho$, the specific enthalpy.

  Non-trivial similarity solutions exist only if the determinant of
system (\ref{sist1})-(\ref{sist5}) vanish. This leads to the condition
\begin{equation}
\label{eq:xi}
  \xi^\pm = \frac{v^x(1 - \omega^2)\pm\omega\sqrt{(1 - v^2)\left[1 -
  v^2\omega^2 - (v^x)^2(1 - \omega^2)\right]}}{1 - v^2\omega^2}
\end{equation}
where $\omega^2 = c_s^2 + v_A^2 - c_s^2v_A^2$ and $\displaystyle{v_A^2
= {b^2}/{\rho \hat{h}}}$ is the Alfv\'en velocity. The plus and minus
signs correspond to rarefaction waves propagating to the left
${\cal R}_{\leftarrow}$ and right ${\cal R}_{\rightarrow}$,
respectively. Note that the values of $\xi$ reduce to those obtained
in PMM by replacing $\omega$ by $c_s$
(i.e., $b = 0$).

  From system (\ref{sist1})-(\ref{sist5}), after some algebraic
manipulations, we get:
\begin{equation}
\label{eqvx}
  W^2\rho \hat{h}(v^x - \xi)\frac{dv^x}{d\xi} + b(1 -
  v^x\xi)\frac{db}{d\xi} + (1 - v^x\xi)\frac{dp}{d\xi} = 0
\end{equation}
\begin{equation}
\label{eqp}
  \frac{dp}{d\xi} = hc_s^2\frac{d\rho}{d\xi}
\end{equation}
\begin{equation}
\label{eqvz}
  \hat{h}Wv^z = {\rm constant}
\end{equation}
\begin{equation}
\label{eqb}
  \frac{b}{\rho} = {\rm constant}.
\end{equation}

  Now, using (\ref{eqp}) and (\ref{eqb}) to eliminate the differentials
of $b$ and $\rho$, and defining $\hat{{\mathcal B}} = b/\rho$, the ODE
(\ref{eqvx}) can be rewritten as
\begin{equation}
\label{eq:dvxdp}
  \displaystyle{
  \frac{dv^x}{dp} = \frac{\left(1 + \frac{\hat{{\mathcal B}}^2
  \rho}{h c_s^2} \right)}{\rho
  \hat{h}W^2}\frac{(1 - \xi v^x)}{(\xi - v^x)}
  },
\end{equation}
in complete analogy with Eq.~(3.20) in Rezzolla, Zanotti \& Pons (2003) 
in the case $b = 0$. The analogy can also be extended 
to the solution procedure. If we define $\hat{{\mathcal A}}=\hat{h}Wv^z$
then, from Eq.~(\ref{eqvz}),
\begin{equation}
\label{eq:vz}
  (v^z)^2 = \hat{{\mathcal A}}^2\left(\frac{1 - (v^x)^2}{\hat{h}^2 +
  \hat{{\mathcal A}}^2}\right),
\end{equation}
which allows to eliminate the dependence on $v^z$ in
(\ref{eq:xi}). Now, from the definition of the Lorentz factor, it can be
derived
\begin{equation}
  W^2=\frac{\hat{h}^2+\hat{{\mathcal A}}^2}{\hat{h}^2(1-(v^x)^2)}
\end{equation}
and, after some algebra, 
\begin{equation}
  \frac{1 - \xi v^x}{\xi - v^x} = \pm \frac{\sqrt{\hat{h}^2 +
  \hat{{\mathcal A}}^2(1 - \omega^2)}}{\hat{h} \omega}
\end{equation}
(where $\omega$ is the positive root of $\omega^2$). Finally,
substituting these last two expressions in (\ref{eq:dvxdp}), we get
\begin{equation}
\label{eq:ode2}
  \displaystyle{
  \frac{dv^x}{1 - (v^x)^2} = \pm \frac{(1 + \frac{\hat{{\mathcal B}}^2
  \rho}{h c_s^2})
  \sqrt{\hat{h}^2 + \hat{{\mathcal A}}^2(1 - \omega^2)}}{\hat{h}^2 +
  \hat{{\mathcal A}}^2} \frac{dp}{\rho \omega}
  }.
\end{equation}
The left hand side of this expression can be integrated analytically
and the right hand side involves only thermodynamical variables and
constants.
Considering that in a Riemann
problem the state ahead of the rarefaction wave is known, the
integration of (\ref{eq:ode2}) allows one to connect the states ahead
($a$) and behind ($b$) the rarefaction wave. The normal velocity
behind the rarefaction wave can be directly obtained as
\begin{equation}
  v_b^x = \tanh \hat{{\mathcal C}},
\end{equation}
where
\begin{equation}
\label{eq:c}
  \hat{{\mathcal C}} = \frac{1}{2} \log \left(\frac{1 + v_a^x}{1 -
  v_a^x}\right) \pm \int_{p_a}^{p_b} \frac{(1 + \frac{\hat{{\mathcal
  B}}^2 \rho}{h c_s^2}) \sqrt{\hat{h}^2 + \hat{{\mathcal A}}^2(1 -
  \omega^2)}}{\hat{h}^2 + \hat{{\mathcal A}}^2} \frac{dp}{\rho \omega}.
\end{equation}
The differential of $p$ in the last integral is taken along the
adiabats of the equation of state. The isentropic character of
rarefaction waves fixes the entropy to that of state $a$,
$s_a$. Having this in mind, the ODE can be integrated, the solution
being only a function of $p_b$. This can be stated in compact form as
\be 
  v^x_b = {\cal R}^a_{\rightleftharpoons}(p_b).  
\ee

  It is interesting to have an expression for the normal
velocity inside the rarefaction wave in terms of the total
pressure. This expression can be built taking into account that, from
the definition of $\hat{p}$, 
\begin{equation}
d\hat{p} = dp + \hat{{\mathcal B}}^2 \rho~d\rho
\end{equation}
and that along an adiabat,
\begin{equation}
dp = h c_s^2 d\rho,
\end{equation}
which combined result in 
\begin{equation}
d\hat{p} = \left( 1 + \frac{\hat{{\mathcal B}}^2 \rho}{h c_s^2} \right) dp.
\end{equation}
Substitution in Eq.~(\ref{eq:c}) gives
\begin{equation}
  \hat{{\mathcal C}} = \frac{1}{2} \log \left(\frac{1 + v_a^x}{1 -
  v_a^x}\right) \pm \int_{\hat{p}_a}^{\hat{p}_b} \frac{\sqrt{\hat{h}^2 +
  \hat{{\mathcal A}}^2(1 - \omega^2)}}{\hat{h}^2 + \hat{{\mathcal A}}^2}
  \frac{d\hat{p}}{\rho \omega}.
\end{equation}
Note that the previous expression is identical to the one derived by
Rezzolla, Zanotti \& Pons (2003) in the case of (non-magnetized)
relativistic hydrodynamics after removing the {\it hats} and
substituting $\omega$ by the sound speed. The corresponding compact
notation for the function giving $v^x_b$ in terms of $\hat{p}_b$ will be
\be 
  v^x_b = \hat{{\cal R}}^{a}_{\rightleftharpoons}(\hat{p}_b).  
\ee

  Function ${\cal R}^a_{\rightleftharpoons}(p)$ is shown in Fig.~\,1,
for different values of the invariant $\hat{{\mathcal B}}$, the
various branches of the curves corresponding to rarefaction waves
propagating towards or away from $a$. Rarefaction waves move towards
(away from) $a$, if the pressure inside the rarefaction is smaller
(larger) than $p_a$. The last assertion also applies for the total
pressure, $\hat{p}$. In a Riemann problem the state $a$ is ahead of
the wave and only those branches corresponding to waves propagating
towards $a$ in Fig.~\,1 must be considered. Moreover, one can
discriminate between waves propagating towards the left and right by
taking into account that the initial left (right) state can only be
reached by a wave propagating towards the left (right). The addition
of a transverse magnetic field in the limiting state forces the value
of the normal velocity within the rarefaction wave to have larger
absolute values. This effect is a consequence of the fact that the
absolute value of the slope of the function ${\cal
R}^a_{\rightleftharpoons}(p)$ at $p_a$, $|{\cal R}^{a \,
'}_{\rightleftharpoons}(p_a)|$, is an increasing function of
$\hat{{\mathcal B}}$ (or $b_a$). Moreover it can be easily proved that
\be
 |{\cal R}^{a \, '}_{\rightleftharpoons}(p_a; \hat{{\mathcal B}}
 \rightarrow \infty)| = \frac{1}{c_{sa}}|{\cal R}^{a \,
   '}_{\rightleftharpoons}(p_a; \hat{{\mathcal B}} = 0)|, 
\ee
where $c_{sa}$ is the sound speed at state $a$. Hence the sound speed
at state $a$ limits in practice the range of values of the normal
velocity within the rarefaction wave (see Fig.~\,1 where the curves
corresponding to $\hat{{\mathcal B}} = 3$ and $10$ almost
coincide). Another consequence of the previous result is that in the
extreme case for which $c_{sa}$ tends to the light speed, the presence
of a transverse magnetic field in the limiting state would have no
practical effect on the rarefaction wave. The effect of the magnetic
field must be combined with the one coming from the presence of
tangential velocities in state $a$ which operates in the opposite
direction (see PMM).

%%%%%%%%%%%%%%%%%%%%%%%%%%%%%%%%%%%%%%%%%%%%%%%%%%%%%%%%%%%%%%%%%%%%
\begin{figure}
\centerline{\psfig{figure=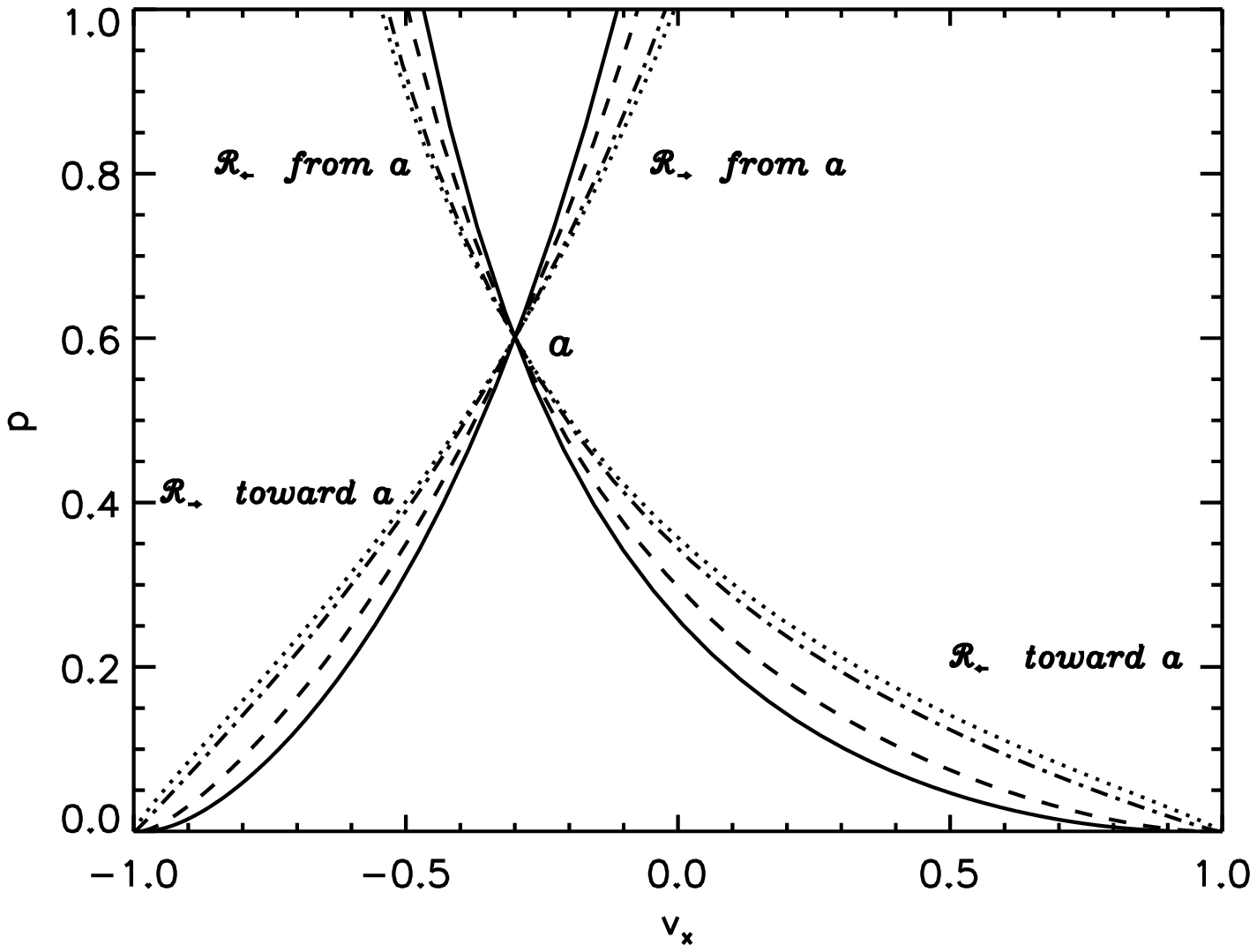,width=14cm}}

{\bf Figure~1}: Loci of states which can be connected with a given
state $a$ by means of relativistic rarefaction waves propagating to
the left (${\cal R}_{\leftarrow}$) and to the right (${\cal
R}_{\rightarrow}$) and moving towards or away from $a$. Solutions for
$\hat{{\mathcal B}} = 0, 1, 3, 10$, correspond to solid, dashed,
dashed-dotted and dotted lines, respectively. The state $a$ is
characterized by $p_a = 0.6$, $\rho_a = 1.0$, and $v^x_a = -0.3$. An
ideal gas EOS with $\gamma = 5/3$ was assumed.
\label{fig1}
\end{figure} 

%
%%%%%%%%%%%%%%%%%%%%%%%%%%%%%%%%%%%%%%%%%%%%%%%%%%%%%%%%%%%%%%%%%%%%%%%%

\section{Jumps across shocks.}

  If $\Sigma$ is a hyper-surface in Minkowski space time across which
$\rho$, $u^{\mu}$, $T^{\mu\nu}$ and $F^{* \mu \nu}$ are discontinuous,
the Rankine-Hugoniot conditions are given by (Lichnerowicz 1967, Anile
1989)
\be 
[\rho u^{\mu}] n_{\mu} = 0 \; ,
\label{rh1}
\ee
\be
[T^{\mu\nu}] n_{\nu} = 0   \; ,
\label{rh2}
\ee
\begin{equation}
\label{RHF}
\left[F^{*\mu\nu}\right]n_\nu =0
\end{equation}
where $n_{\mu}$ is the unit normal to $\Sigma$, and where we have used the
notation
\be
[G] = G_a-G_b,
\ee
$G_a$ and $G_b$ being the boundary values of $G$ on the two sides of
$\Sigma$.

  Considering $\Sigma$ as the hyper-surface in four-dimensional space
describing the evolution of a shock wave normal to the $x$-axis, the
unitarity of $n_{\nu}$ allows one to write it as
\be
n^{\nu} = W_s (V_s,1,0,0),
\ee 
where $V_s$ is interpreted as the coordinate velocity of the surface
that defines the position of the shock wave and $W_s$ is the Lorentz
factor of the shock,
\be
W_s = \frac{1}{\sqrt{1-V_s^2}}.
\ee

  Equations (\ref{rh1}) and (\ref{RHF}) allow one to introduce two
invariants across the shock
\be
j \equiv W_s D_a (V_s-v^x_a) = W_s D_b (V_s-v^x_b),
\label{mflux}
\ee
\be
f \equiv W_s B_a (V_s-v^x_a) = W_s B_b (V_s-v^x_b),
\label{bflux}
\ee
where $B = bW$. Quantity $j$ represents the mass flux across the shock
and according to our definition, $j$ is positive for shocks
propagating to the right (the same convention as the one used in
Mart\'{\i} \& M\"uller 1994, and PMM). 
Dividing Eq.~(\ref{mflux}) by Eq.~(\ref{bflux}), we get that
the quantity $B/D$ (or, equivalently, $b/\rho$) is constant across
shocks as it was through rarefaction waves.
                                              
  Next, the Rankine-Hugoniot conditions (\ref{rh1}), (\ref{rh2}) can
be written in terms of the conserved quantities $D$, $\hat{S}^j$ and
$\hat{\tau}$, and $j$ as follows
\begin{equation}
  \label{ch1}
  \left[v^x\right]=-\frac{j}{W_s}\left[\frac{1}{D}\right]
\end{equation}
\begin{equation}
  \label{ch2}
  \left[\hat{p}\right]=\frac{j}{W_s}\left[\frac{\hat{S}^x}{D}\right]
\end{equation}
\begin{equation}
  \label{ch3}
  \left[\frac{\hat{S}^z}{D}\right]=0
\end{equation}
\begin{equation}
  \label{ch4}
  \left[v^x\hat{p}\right]=\frac{j}{W_s}\left[\frac{\hat{\tau}}{D}\right].
\end{equation}

  Now from Eq.~(\ref{ch3}) we have that the quantity $\hat{h}Wv^z$ is
constant across the shock, as it is through rarefactions. 

  We note that in deriving equations (\ref{ch1})--(\ref{ch4}) we have
made use of the fact that the mass flux is nonzero across a shock.
The conditions across a tangential discontinuity imply continuous
total pressure and normal velocity (by setting $j=0$ in equations
(\ref{ch1}), (\ref{ch2}) and (\ref{ch4})), and an arbitrary jump in the
tangential velocity and transverse magnetic field. 

  Our aim now is to write $v_b^x$, the normal flow speed in the
post-shock state, as a function of the post-shock pressure $p_b$. As a
first step, we write $v_b^x$ as a function of $p_b^*$, $j$ and $V_s$
(and the preshock state, $a$). Given the complete analogy between the
jump conditions Eqs.~(\ref{mflux}), (\ref{ch1})-(\ref{ch4}), and the
corresponding expressions in PMM, we write
\be 
  v^x_b = \left( \hat{h}_a W_a v^x_a + \frac{W_s
  (\hat{p}_b-\hat{p}_a)}{j} \right)
  \left( \hat{h}_a W_a + (\hat{p}_b - \hat{p}_a) 
  \left(\frac{W_s v^x_a}{j} + \frac{1}{\rho_a W_a} \right) 
  \right)^{-1}.  
\label{eq:vxbshock}
\ee
The dependence on the magnetic
field in the pre- and post-shock states is hidden in the definitions
of $\hat{p}$ and $\hat{h}$.  

  The shock speed $V_s$ can be eliminated using the definition of mass
flux to obtain
\begin{equation}
\displaystyle{
V_s^{\pm} = \frac{ \rho_a^2 W_a^2 v^x_a  \pm   
                   |j| \sqrt{j^2 + \rho_a^2 W_a^2 (1 -{v_a^x}^2)}
            }{ \rho_a^2 W_a^2 + j^2 }
}
\label{velshock}
\end{equation}
where $V_s^{+}$ ($V_s^{-}$) corresponds to shocks propagating to the
right (left). 

  Proceeding in the same way as in PMM (i.e.,
$\left[T^{\mu\nu}\right]n_\nu\lbrace (\hat{h}u_\mu)_a +
(\hat{h}u_\mu)_b\rbrace = 0$) to derive the Taub's adiabat, we can now obtain
\begin{equation}
  \left[\hat{h}^2\right] = \left(\frac{\hat{h}_b}{\rho_b} +
  \frac{\hat{h}_a}{\rho_a}\right)\left[\hat{p}\right],
\label{eq:lich}
\end{equation}
i.e, the Lichnerowicz's adiabat (Anile 1989) particularized to our special
setup.  Figure~2 represents the function
\begin{equation}
\displaystyle{{\mathcal L}^a(\hat{h};\hat{p}_b) \equiv
\hat{h}^{2} - \hat{h}_a^{2} - \left(\frac{\hat{h}}{\rho(\hat{h},\hat{p}_b)} +
\frac{\hat{h}_a}{\rho_a}\right)(\hat{p}_b - \hat{p}_a)}
\label{eq:la}
\end{equation}
for an ideal gas equation of state although the general shape of the
curve (positive asymptotic branches; negative value for $\hat{h}=1$)
is independent of the equation of state. The (unique) root at the
right of $\hat{h}=1$ defines the thermodynamical post-shock state.

%%%%%%%%%%%%%%%%%%%%%%%%%%%%%%%%%%%%%%%%%%%%%%%%%%%%%%%%%%%%%%%%%%%%%%
%
\begin{figure}
\centerline{\psfig{figure=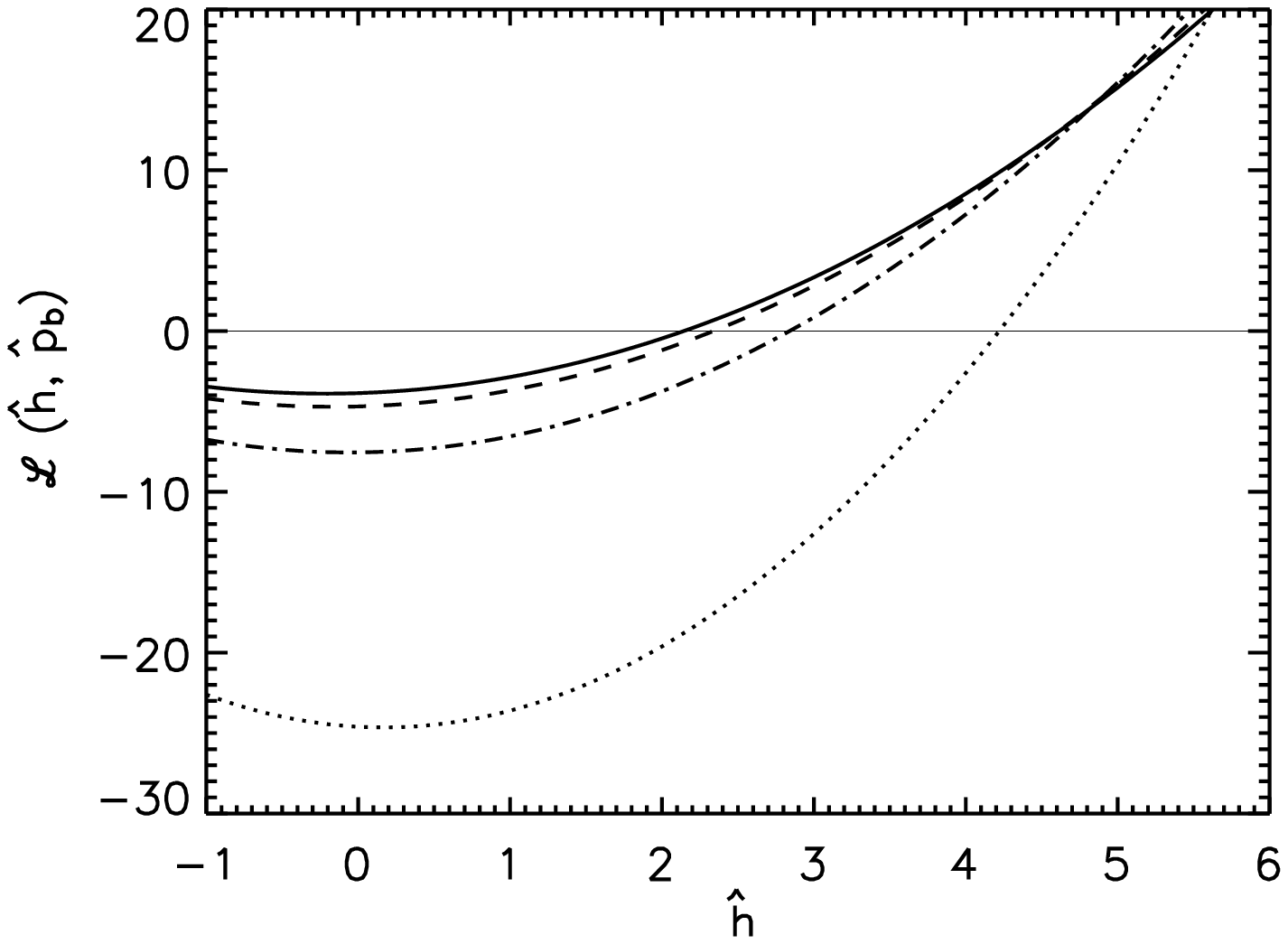,width=14cm}}

{\bf Figure~2}: Graphical representation of the function ${\mathcal
L}^a(\hat{h};\hat{p}_b)$ (definition: see text) whose zeroes for
varying post-shock pressures, $\hat{p}_b$, define the Lichnerowicz's
adiabat. The state $a$ is characterized by $p_a = 0.25$, $\rho_a =
1.0$ and $v^x_a =0.5$. Solutions for transversal magnetic fields in
state $a$, $\hat{{\mathcal B}}_a = 0, 0.5, 1.0, 2.0$, correspond to
solid, dashed, dashed-dotted and dotted lines, respectively.
$\hat{p}_b$ was chosen to be 1.0.  An ideal gas EOS with $\gamma =
5/3$ was assumed.
\label{fig2}
\end{figure} 
%
%%%%%%%%%%%%%%%%%%%%%%%%%%%%%%%%%%%%%%%%%%%%%%%%%%%%%%%%%%%%%%%%%%%%%%%

Equation~(\ref{eq:lich}) together with the definitions of $\hat{p}$
and $\hat{h}$, the equation of state and the constancy of $b/\rho$
through the shock, allows to write $\rho_b$ as a function of $p_b$ and
the preshock state $a$. Next, multiplying (\ref{rh2}) by $n_\mu$ and
using the definition of relativistic mass flux one obtains
\begin{equation}
\label{j}
  j^2=\frac{-\left[\hat{p}\right]}{\left[\hat{h}/\rho\right]}.
\end{equation}

  Using the positive (negative) root of $j^2$ for shock waves
propagating towards the right (left), equation (\ref{j}) allows one to
obtain the desired relation between the post-shock normal velocity
$v^x_b$ and the post-shock pressure $p_b$. In a compact way the
relation reads
\be 
  v^x_b = {\cal S}^a_{\rightleftharpoons}(p_b).  
\ee
Alternatively, the relation can be written as a function of $\hat{p}_b$
\be 
  v^x_b = \hat{{\cal S}}^{a}_{\rightleftharpoons}(\hat{p}_b).  
\ee

  Let us note that the expressions used to build up the function
$\hat{{\cal S}}^{a}_{\rightleftharpoons}$, namely
Eqs.~(\ref{eq:vxbshock}), (\ref{velshock}), (\ref{eq:lich}) and
(\ref{j}), are formally identical to those corresponding to the pure
(i.e., non-magnetized) relativistic hydrodynamical case. The
difference appears in the definition of the function $\rho =
\rho(\hat{h}, \hat{p})$, which leads to different roots of the
function ${\mathcal L}^a(\hat{h};\hat{p}_b)$, Eq.~(\ref{eq:la}).
Function ${\cal S}^a_{\rightleftharpoons}(p)$ is shown in Fig.~\,3,
for different values of the invariant $\hat{{\mathcal B}}$, the various
branches of the curves corresponding to shock waves propagating
towards or away from $a$.  In order to select the relevant branch of
the function ${\cal S}^a_{\rightleftharpoons}(p)$ the same
argumentation as in the case of rarefaction waves can be used (see
\S3). As in the case of rarefaction waves, the addition of a
transverse magnetic field in the limiting state forces the value of
the normal velocity in the pre/post shock state to have larger
absolute values.  Again, this effect must be combined with the one
coming from the presence of tangential velocities in state $a$ (see
PMM).

%%%%%%%%%%%%%%%%%%%%%%%%%%%%%%%%%%%%%%%%%%%%%%%%%%%%%%%%%%%%%%%%%%%%%%%
%
\begin{figure}
\centerline{\psfig{figure=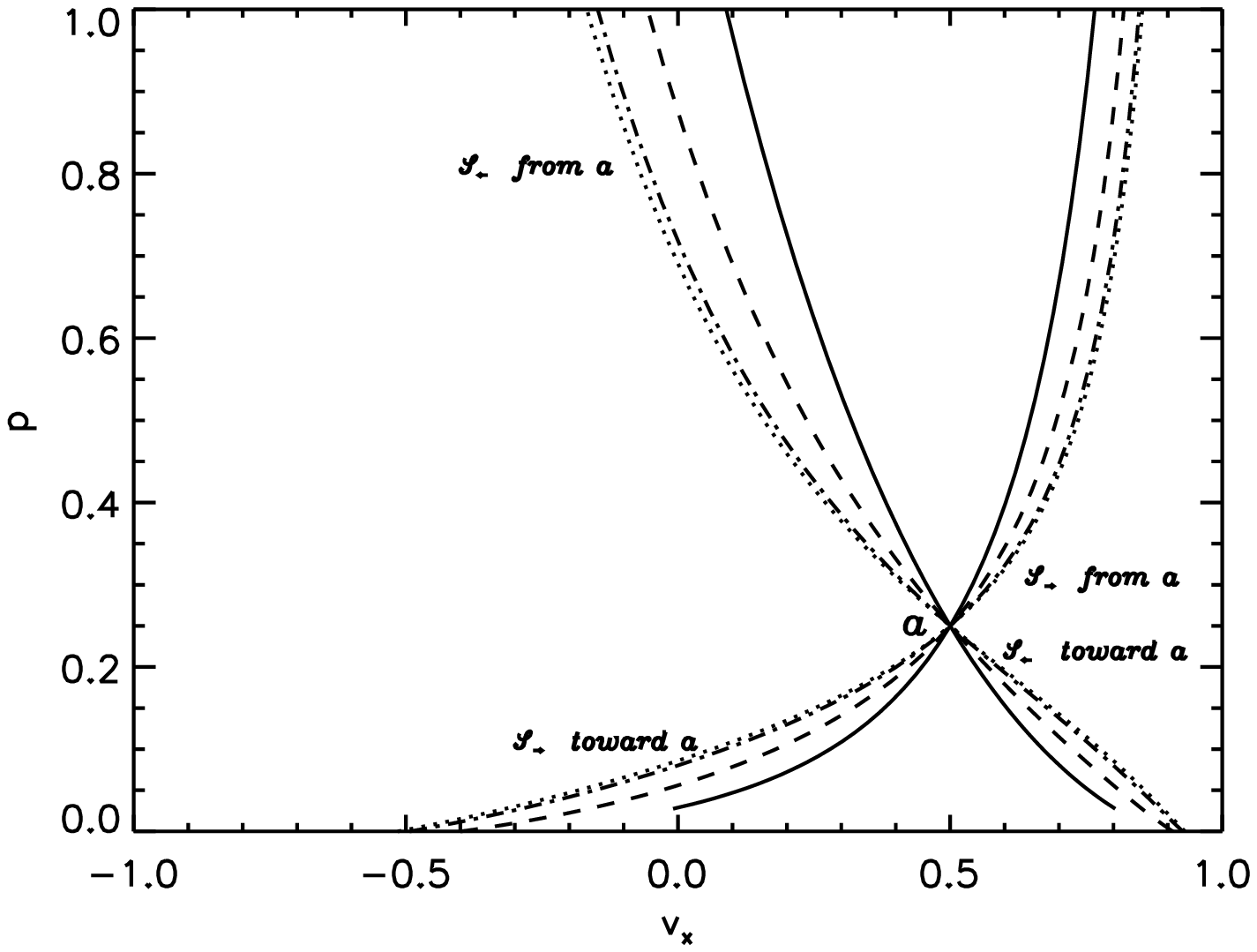,width=14cm}}

{\bf Figure~3}: Loci of states which can be connected with a given
state $a$ by means of relativistic shock waves propagating to the left
(${\cal S}_{\leftarrow}$) and to the right (${\cal S}_{\rightarrow}$)
and moving towards or away from $a$. Solutions for $\hat{{\mathcal B}}
= 0, 1, 3, 10$, correspond to solid, dashed, dashed-dotted and dotted
lines, respectively. The state $a$ is characterized by $p_a = 0.25$,
$\rho_a = 1.0$, and $v^x_a = 0.5$. An ideal gas EOS with $\gamma =
5/3$ was assumed.
\label{fig3}
\end{figure} 
%
%%%%%%%%%%%%%%%%%%%%%%%%%%%%%%%%%%%%%%%%%%%%%%%%%%%%%%%%%%%%%%%%%%%%%%%

  Once $v^x_b$ is know, $v^z_b$ can be obtained through 
\begin{equation}
\label{eq:vz_2}
  (v^z_b)^2 = \hat{{\mathcal A}}^2\left(\frac{1 - (v^x_b)^2}{\hat{h}_b^2 +
  \hat{{\mathcal A}}^2}\right),
\end{equation}
where we have defined $\hat{{\mathcal A}}=\hat{h}_aW_av^z_a$. Analogously, 
$b_b = \hat{{\mathcal B}} \rho_b$
with $\hat{{\mathcal B}} = b_a/\rho_a$.

%%%%%%%%%%%%%%%%%%%%%%%%%%%%%%%%%%%%%%%%%%%%%%%%%%%%%%5
\section{The solution of the Riemann problem.}

  As discussed in \S~2, for the particular case under consideration
(magnetic field orthogonal to both the fluid velocity and the wave
propagation direction), the time evolution of a Riemann problem with
initial states $L$ (left) and $R$ (right) can be represented as:
\be LR\; \rightarrow \;L\;{\cal W}_{\leftarrow}\;L_*\;{\cal C}\;
  R_*\;{\cal W}_{\rightarrow}\;R 
\ee 
where $\cal W$ and $\cal C$ denote a (fast magnetosonic-) shock or
rarefaction, and a contact discontinuity, respectively.  The arrows
($\leftarrow$ / $\rightarrow$) indicate the direction (left / right)
from which fluid elements enter the corresponding wave.

  The solution of the Riemann problem consists in finding the
intermediate states, $L_*$ and $R_*$, as well as the positions of the
waves separating the four states (which only depend on $L$, $L_*$,
$R_*$ and $R$). The functions ${\cal W}_{\rightarrow}$ and ${\cal
W}_{\leftarrow}$ allow one to determine the functions $v^x_{R*}(\hat{p})$
and $v^x_{L*}(\hat{p})$, respectively. The pressure $\hat{p}_*$ and the flow
velocity $v^x_*$ in the intermediate states are then given by the
condition
\be
v^x_{R*}(\hat{p}_*) = v^x_{L*}(\hat{p}_*) = v^x_*.
\ee

  The functions $v^x_{S*}(\hat{p})$ are defined by
\begin{equation}
v^x_{S*}(\hat{p}) = \left\{ \begin{array}{ll} 
                    \hat{{\cal
                        R}}^S(\hat{p}) & \mbox{if $\hat{p} \leq
                        \hat{p}_S$} \\ 
                    \hat{{\cal S}}^S(\hat{p}) & \mbox{if $\hat{p} >
                        \hat{p}_S$ \, ,}
                        \end{array}
                        \right.
\end{equation}
where $\hat{{\cal R}}^S(\hat{p})$ ($\hat{{\cal S}}^S(\hat{p})$)
denotes the family of all states which can be connected through a
rarefaction (shock) with a given state $S$ ($L,R$) ahead of the
wave. Once $\hat{p}_*$ and $v^x_*$ have been obtained the remaining
quantities can be computed.

  Figure~\,4 shows the solution of a particular Riemann problem for
different values of the magnetic field $b = 0, 1.0, 2.0, 4.0$ in the
initial states. The crossing point of any two lines gives the pressure
and the normal velocity in the intermediate states. 

%%%%%%%%%%%%%%%%%%%%%%%%%%%%%%%%%%%%%%%%%%%%%%%%%%%%%%%%%%%%%%%%%%%%%
%
\begin{figure}
\centerline{\psfig{figure=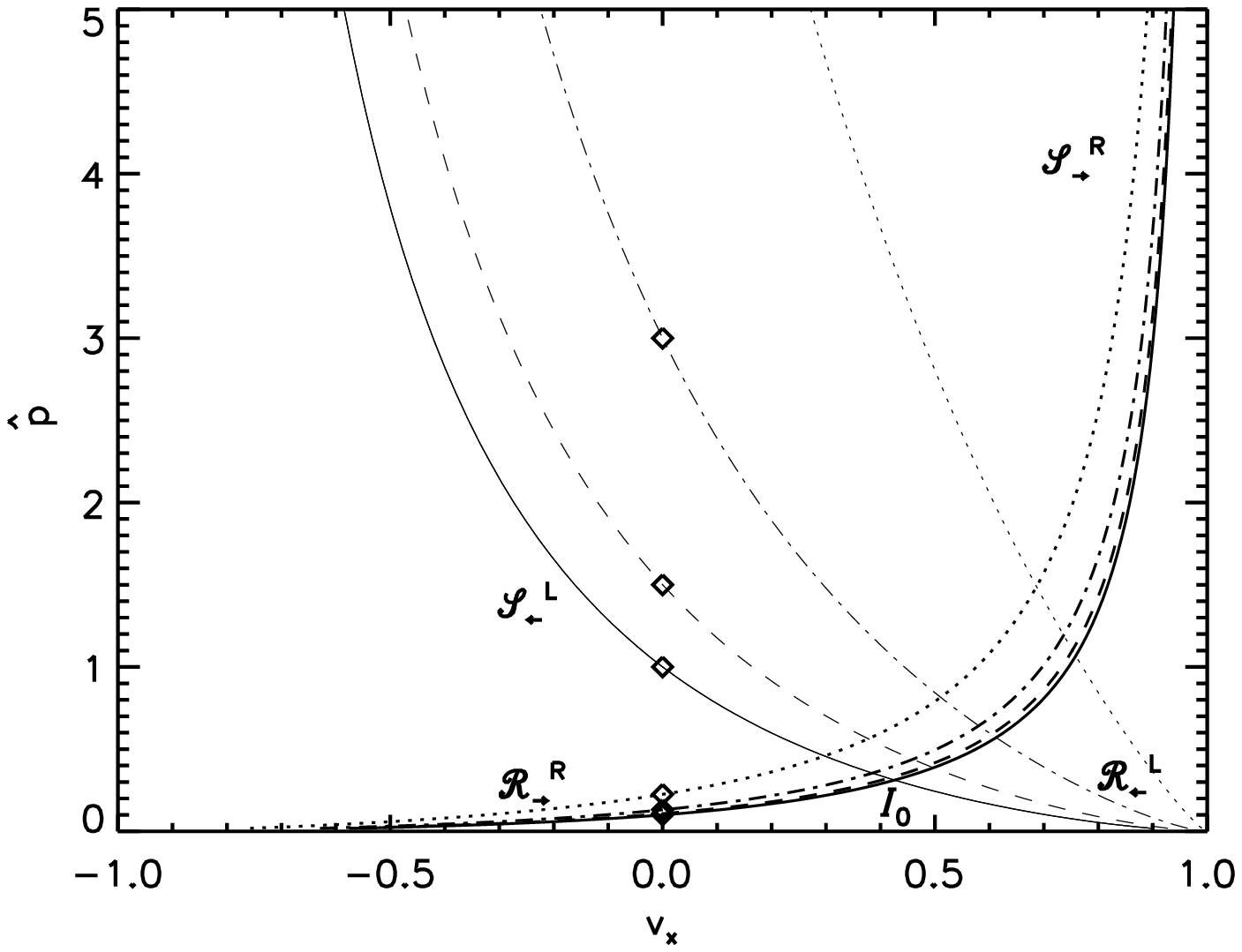,width=14cm}}
\label{fig4}

{\bf Figure~4}: Graphical solution in the $(\hat{p},v^x)$--plane of
the relativistic Riemann problem with initial data $p_L=1.0$,
$\rho_L=1.0$, $v^x_L=0.0$; $p_R=0.1$, $\rho_R=0.125$ and $v^x_R=0.0$
for different values of the transversal magnetic field $\hat{\cal B} =
0, 1.0, 2.0, 4.0$, in the left and right states represented by solid,
dashed, dashed-dotted and dotted lines, respectively. Diamonds
indicate the initial states. An ideal gas EOS with $\gamma = 1.4$ was
assumed.  The crossing point of any two lines gives the pressure and
the normal velocity in the intermediate states.  $I_0$ gives the
solution for vanishing magnetic field.

\end{figure}
%
%%%%%%%%%%%%%%%%%%%%%%%%%%%%%%%%%%%%%%%%%%%%%%%%%%%%%%%%%%%%%%%%%%%%%%%%

  It must be noted that the resolution of the Riemann problem under
consideration can be formally done in the same way as the pure
(relativistic) hydrodynamical problem with some modifications (see the
Appendix).  First of all, $p$ and $h$ have to be replaced by $\hat{p}$
and $\hat{h}$. Secondly, the sound speed in the integrand of
Eq.~(\ref{eq:c}) has to be replaced by $\omega$.  Finally, the
equation of state that provides the rest-mass density as a function of
the pressure and the enthalpy, $\rho(\hat{h}, \hat{p})$, has to be
modified to include the contributions from the magnetic field. Given
the parallelism between the present particular (relativistic)
magnetohydrodynamical case and the purely hydrodynamical one, the
effects concerning the smooth transition from one wave pattern to
another when the tangential velocities in the initial states are
changed (Rezzolla \& Zanotti 2002, Rezzolla, Zanotti \& Pons 2003)
will extend to the present case for fixed initial values of the
magnetic field. Hence we concentrate in the effects on the solution
induced by varying the initial magnetic fields.  Figure~\,5 shows the
solution of a Riemann problem with a) vanishing magnetic field, and b)
$b_R=0.8$. Whereas in the purely hydrodynamical case the Riemann
solution gives rise to a left propagating rarefaction wave and right
propagating contact and shock waves, the case with non vanishing
magnetic field leads to a couple of shock waves and a left propagating
contact discontinuity. The reason for this qualitative change
(rarefaction/shock to shock/shock) can be found in the increase of the
total pressure in the initial right state of the magnetized case.
This increases the total pressure in the intermediate states at
the two sides of the contact discontinuity. When this pressure becomes
larger than that at the initial left state then a left propagating
shock instead a rarefaction is produced.

  Also noticeable from Fig.~\,5 is the increase of velocity of the
shock propagating towards the right in the magnetized case.  The
increase of the velocity of propagation of fast waves for increasing
magnetic fields (approaching the light speed in the fluid rest frame
for strong enough fields, much larger than equipartition) is
well-known in RMHD (e.g., Anile 1989). In the particular magnetic
problem under consideration it leads to the increase of velocity of
propagation of rarefaction heads and tails, and of shocks and can be
understood as follows. Equation (\ref{eq:xi}) gives the propagation
speed of the head/tail of a right/left propagating rarefaction wave on
a given state.  Taking $v^x, v = 0$ in Eq.~(\ref{eq:xi}), one gets
$|\xi^\pm| = \omega$ and $\omega \rightarrow 1$ for large enough
$b$. One should remember that for purely hydrodynamical (relativistic)
flows, $|\xi^\pm|$ has the sound speed as limiting value. A similar
result holds for shocks. For preshock states at rest, $|V^\pm_s|
\rightarrow 1$ as the magnetic field in the preshock state is
increased.  This can be seen by remembering that the shocks that
appear in our configurations are super-magnetosonic, and that the fast
magnetosonic speed in the preshock state ($\omega_a$) tends to light
speed when $b_a \rightarrow \infty$.

  Finally, let us note that the drift towards solutions involving
only discontinuous waves (shock waves and rarefactions of negligible
width) for increasing magnetic fields as concluded in the previous
paragraph, is consistent with the fact that in the limit of strong
magnetization, the equations of RMHD reduce to the equations of
force-free degenerate electrodynamics, whose Riemann problem only
involves (linearly degenerate) discontinuous waves (Komissarov 2002).

%%%%%%%%%%%%%%%%%%%%%%%%%%%%%%%%%%%%%%%%%%%%%%%%%%%%%%%%%%%%%%%%%%%%%%
%
\begin{figure}
\centerline{\psfig{figure=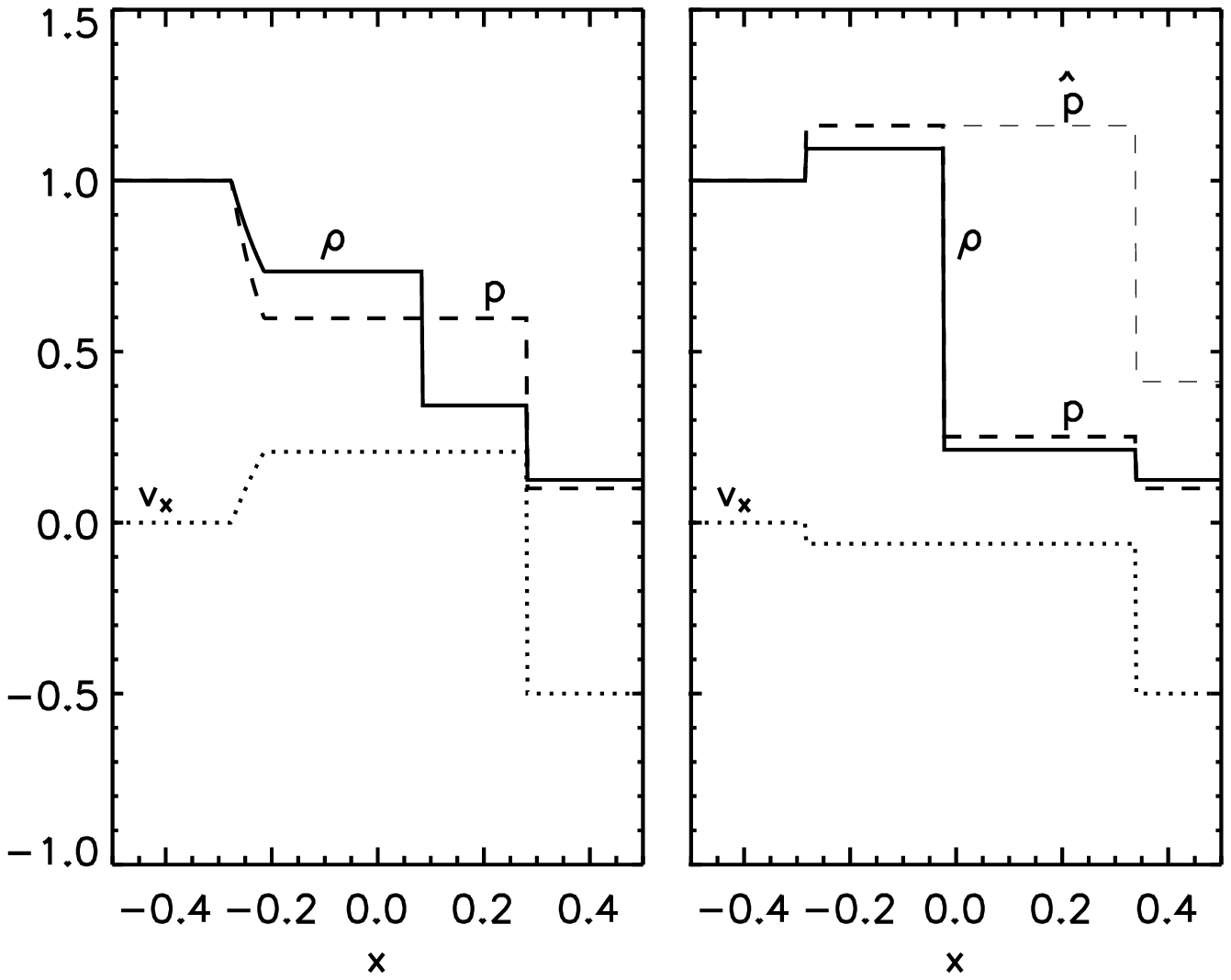,width=14cm}}
\label{fig5}

{\bf Figure~5}: Solution of the Riemann problem with initial data
$p_L=1.0$, $\rho_L=1.0$, $v^x_L=0.0$; $p_R=0.1$, $\rho_R=0.125$ and
$v^x_R=-0.5$, at time $t = 0.4$. An ideal gas EOS with $\gamma=5/3$
was assumed. Left panel: purely hydrodynamical problem. Right panel:
with $b_R = 1.0$. Note that ${\hat p}$ is constant through the contact
discontinuity whereas the thermal pressure, $p$, is not.

\end{figure}
%
%%%%%%%%%%%%%%%%%%%%%%%%%%%%%%%%%%%%%%%%%%%%%%%%%%%%%%%%%%%%%%%%%%%%%%%%

%%%%%%%%%%%%%%%%%%%%%%%%%%%%%%%%%%%%%%%%%%%%%%%%%%%%%%%%%%%%%%%%%%%%%%%%%%%

\section{Summary and conclusions}

  We have obtained an exact solution of the Riemann problem for
multidimensional relativistic magnetohydrodynamics in the particular
case in which the magnetic field is normal to the fluid velocity. In
this particular problem, the complex 7--wave pattern of RMHD is
reduced to two fast magnetosonic waves and a contact discontinuity,
which allows to use the same procedure as in the non magnetic
case. Alternatively, we have shown that the problem can be understood
as a purely RHD situation with a modified equation of state (see
Appendix A for details).

  Two interesting features arise from our results. First, for fixed
initial thermodynamical states, it is possible to change continuously
from one wave pattern to another (shock/shock, shock/rarefaction,
rarefaction/rarefaction) analogously to what happens when tangential
velocities are introduced (Rezzolla \& Zanotti 2002). Secondly,
we recover the result for RMHD flows with general magnetic field
configurations stablishing the tendency of the wave speeds to the
light speed when the magnetic field dominates the thermodynamical
pressure and energy. For our particular configuration of the magnetic
field, this results in fast moving shock waves and rarefaction waves
in which the distance between the head and the tail is progressively
reduced as the magnetic field increases. The drift towards
solutions involving only discontinuous waves (shock waves and
rarefactions of negligible width) for increasing magnetic fields is
consistent with the fact that in the limit of strong magnetization,
the equations of RMHD reduce to the equations of force-free degenerate
electrodynamics, whose Riemann problem only involves discontinuous
waves (Komissarov 2002).

  In addition to the theoretical interest of our results by their own,
having obtained an exact solution of the RMHD Riemann problem is
relevant for the development of numerical codes. Up to now, in order
to test the various algorithms and approximate Riemann solvers
developed for numerical applications, one could only increase the
spatial resolution and hope that the numerical solution converged to
the physical one.  Having an exact solution to compare with, even if
it is just a particular case, allows for a more rigorous testing and
error estimation. Last but not least, from a pedagogical point of
view, it is more convenient to start understanding and solving a
simpler case before attempting the solution of the full problem, which
is the next natural extension of this work.

\bigskip
%== =====================================================
\acknowledgements
%=======================================================

It is a pleasure to thank L. Ant\'on and L. Rezzolla for useful
discussions and comments. Financial support for this research has been
provided by Spanish MEC grant AYA2004-08067-C03. J.A.P. is supported
by a {\it Ram\'on y Cajal} contract.

%%%%%%%%%%%%%%%%%%%%%%%%%%%%%%%%%%%%%%%%%%%%%%%%%%%%%%%%%%%%%%%%%%%%%%%%%%%%%%
\appendix

\section{A hydrodynamical approach}

  Equations (\ref{eq:cl}) are identical to those for the purely
hydrodynamical case by replacing
\begin{equation}
  p \longrightarrow \hat{p} = p + \frac{b^2}{2}
\end{equation}
\begin{equation}
  h \longrightarrow \hat{h} = h + \frac{b^2}{\rho},
\end{equation}
indicating that a description of the present particular RMHD problem
based on a purely hydrodynamical approach with a different equation of
state may be possible. In this Appendix, we explore such a
possibility, first suggested by Romero et al. (1996). The key point is
to eliminate the magnetic field from the equations by building up a
thermodynamically consistent equation of state including the effects
of the magnetic field.

  It follows from eqs.~(\ref{eq:cl}) that
\begin{equation}
\frac{D(b/\rho)}{Dt} = 0,
\end{equation} 
where $D\,\,/Dt$ stands for the standard convective derivative,
implying that the evolution of the fluid elements is along states
keeping $b/\rho = $ constant. Then for a particular fluid element,
$\hat{p}$ and $\hat{h}$ can be written as
\begin{equation}
  \hat{p} = p + \hat{{\mathcal B}}^2 \rho^2/2 
\label{eq:p*}
\end{equation}
\begin{equation}
  \hat{h} = 1 + \varepsilon + \frac{\hat{p}}{\rho} + \hat{{\mathcal
  B}}^2 \rho/2,
\end{equation}
where $\hat{{\mathcal B}}$ is a constant.

Consistency of the two previous expressions with the definition of
$\hat{h}$ is fulfilled by defining a new specific internal energy,
$\hat{\varepsilon}$,
\begin{equation}
  \hat{\varepsilon} = \varepsilon + \hat{{\mathcal B}}^2 \rho/2.
\label{eq:epsi*}
\end{equation}

  The evolution of the fluid elements in a perfect fluid is
adiabatic. Hence now we look for the adiabats of the new equation of
state, $\hat{p} = \hat{p}(\rho,\hat{\varepsilon})$. The fact that the
evolution of the fluid elements keeps $\hat{{\mathcal B}}$ = constant
draw us to consider that $\hat{{\mathcal B}}$ is constant along the
adiabats of the new equation of state, $\hat{s} = $ constant. We shall
use this fact to look for the desired relation between the entropies
of the two equations of state, $s$ and $\hat{s}$. To do this, we
differentiate eq.~(\ref{eq:epsi*}) along transformations keeping
$\hat{s} = $ constant. We get
\begin{equation}
  d\hat{\varepsilon} = d\varepsilon + \hat{{\mathcal B}}^2 d\rho/2.
\label{eq:depsi*}
\end{equation}
On the other hand, according to the first law of thermodynamics, for
an adiabatic transformation,
\begin{equation}
  d\hat{\varepsilon} = \frac{\hat{p}}{\rho^2} d\rho.
\end{equation}
Substitution of $\hat{p}$ in the previous expression gives 
\begin{equation}
  d\hat{\varepsilon} = \frac{p}{\rho^2} d\rho + \hat{{\mathcal B}}^2 d\rho/2.
\end{equation}
Finally, comparison with eq.~(\ref{eq:depsi*}) leads to
\begin{equation}
  d\varepsilon = \frac{p}{\rho^2} d\rho,
\end{equation}
which is formally identical to the variation of internal energy in an
adiabatic transformation $s = $ constant. Taking into account that the
differentials were taken along the adiabats of the new equation of
state, the conclusion is that the entropy in the {\it new} equation of
state must be a function of the entropy in the original equation of
state only.

  Now the sound speed of the new equation of state, 
\begin{equation}
\hat{c}_s = \sqrt{\frac{1}{\hat{h}} \left.\frac{\partial \hat{p}}{\partial
\rho}\right|_{\hat{s}}}
\end{equation}
can be derived. The substitution of $\hat{p}$ following eq.~(\ref{eq:p*})
and the equivalence of the adiabats of the two equations of state
leads to 
\begin{equation}
\hat{c}_s = \sqrt{\frac{h}{\hat{h}} c_s^2 + \frac{\hat{{\cal B}}^2
\rho}{\hat{h}}},
\end{equation}
where $h$ and $c_s$ stand for the enthalpy and the sound speed of
the original equation of state, respectively. Finally, a bit of
algebra allows us to write,
\begin{equation}
\hat{c}_s = \sqrt{(1- v_A^2) c_s^2 + v_A^2},
\end{equation}
where $v_A$ is the Alfven speed for our particular case in which only
one component of the magnetic field is non-zero, $\displaystyle{v_A =
b/\sqrt{\rho \hat{h}}}$ $\left(\displaystyle{ = \hat{{\mathcal B}}
  \sqrt{\frac{\rho}{h+\hat{{\mathcal B}}^2 \rho}}}\right)$. Note that
$\hat{c}_s$ coincides with the quantity $\omega$ defined in \S~2.

  It is interesting to note that although the original equation of
state could have a sound speed significantly smaller than light speed
(e.g., $\leq 1/\sqrt{3}$, for an ultra-relativistic non-degenerate
ideal gas), the sound speed of the new equation of state (that
represents the true propagation speed of perturbations in our
magnetized fluid) approaches the light speed for large enough values
of $\hat{{\mathcal B}}$ (or $b$).  Finally, notice that the convexity
of the EOS is ensured, since ${\mathcal B}^2$ is a positive defined
quantity and therefore ${\displaystyle
  \left. \frac{\partial^2 \hat{p}}{\partial\rho^2}\right|_s >0 }$.

%%%%%%%%%%%%%%%%%%%%%%%%%%%%%%%%%%%%%%%%%%%%%%%%%%%%%%%%%%%%%%%%%%%%%%%%%%%%%

\end{document}